# Complex Quadrature Spatial Modulation

Manar Mohaisen and Saetbyeol Lee

In this paper, we propose a spatial modulation (SM) scheme referred to as complex quadrature spatial modulation (CQSM). In contrast to quadrature spatial modulation (QSM), CQSM transmits two complex signal constellation symbols on the real and quadrature spatial dimensions at each channel use, increasing the spectral efficiency. To this end, signal symbols transmitted at any given time instant are drawn from two different modulation sets. The first modulation set is any of the conventional QAM/PSK alphabets, while the second is a rotated version of it. The optimal rotation angle is obtained through simulations for several modulation schemes and analytically proven for the case of QPSK, where both results coincide. Simulation results showed that CQSM outperformed QSM and generalized SM (GSM) by approximately 5 and 4.5 dB, respectively, for the same transmission rate. Its performance was similar to that of QSM; however, it achieved higher transmission rates. It was additionally shown numerically and analytically that CQSM outperformed QSM for a relatively large number of transmit antennas.

## I. Introduction

Spatial modulation (SM) has emerged in the last decade as a multiple-input multiple-output (MIMO) technique that utilizes both signal constellation symbols, such as the quadrature amplitude and phase shift keying modulation (QAM/PSK) alphabet, and the spatial dimension, which is the index of a single or multiple transmit antennas, to convey information from transmitter to receiver [1]. The fundamental concept of SM is that a transmitter equipped with a single radio frequency (RF) chain and several physical antennas can have a relatively high capacity through using the spatial dimension, represented by the indices of the transmit antennas, to convey information to the receiver. The receiver then recovers the designated information through demodulating both the signal symbol and transmit antenna index.

A special case of the SM is referred to as space shift keying (SSK), whereby the conventional modulated symbols, QAM/PSK, are replaced by the presence or absence of energy assigned to a particular antenna [2]. Both SM and SSK schemes were generalized in [3] and [4], where more than one transmit antenna can be simultaneously used to increase the system's spectral efficiency or reduce the number of required physical antennas to achieve a target performance criterion.

A generalized SM scheme for large-scale MIMO systems was proposed in [5]. An improved SM (ISM) was proposed in [6]–[7], where a simple mapping from the input bits to the transmission vectors is performed for any number of transmit antennas. Hence, the log-two number of the transmit antenna condition imposed in the conventional SM and SSK schemes is relaxed. Moreover, a combination of space-time block coding and the SM scheme is introduced in [8], where the proposed technique outperformed the conventional SM in terms of error performance by 3 to 5 dB, while achieving the same spectral efficiency. In [9], a precoding-aided spatial modulation (PSM) approach is proposed in which the transmitted symbol is precoded using a channel matrix-based criterion so that a single receive antenna is activated, thereby conveying additional information to the receiver. In addition, antenna selection techniques for both SM and PSK schemes have been proposed to achieve further diversity and power gains (see [10] and [11] and their references). Detailed comparisons among spatial modulation schemes are introduced in [12] and [13].

Recently, a quadrature spatial modulation (QSM) scheme was introduced in [14]. In QSM, the spatial constellation symbols are expanded into in-phase and quadrature dimensions. The real part of the signal constellation symbol is transmitted on the first spatial constellation dimension; the imaginary part is transmitted on the second. Since real and imaginary parts of the signal constellation symbol are transmitted over orthogonal carriers, QSM does not suffer inter-channel interference (ICI). Based on its structure, QSM increases the spectral efficiency by $\log_2(n_T)$ bits/s/Hz compared to the conventional SM, which achieves $q + \log_2(n_T)$ bits/s/Hz, where $q$ and $n_T$ denote the number of bits per signal constellation symbol and number of physical antennas, respectively. The QSM scheme is used in [15] to efficiently mitigate eavesdropping. Additionally, a precoding-aided QSM is introduced in [16], where the indexes of the designated receive antennas are used to convey information.

In this paper, we advance the conventional SM technique to achieve a spectral efficiency of $2(q + \log_2(n_T))$ bits/s/Hz. The proposed scheme is called complex quadrature spatial modulation (CQSM). Instead of transmitting the real and imaginary parts of a signal constellation symbol on a designated spatial constellation dimension, CQSM transmits two complex signal modulation symbols, drawn from two different modulation sets, at each channel use. The first and second symbols are drawn from a conventional PSK/QAM modulation set and a rotated version of it, respectively. The rotation angle has a direct impact on the bit-error-rate (BER) performance of the CQSM scheme. Therefore, after introducing CQSM, the rotation angle is optimized through extensive Monte Carlo simulations. The optimal value is analytically obtained in Appendix I for the case of QPSK modulation. It is herein shown, both numerically and analytically, that CQSM outperforms QSM for a high number of transmit antennas, while achieving a higher spectral efficiency. For the same spectral efficiency, CQSM outperforms SM, generalized SM (GSM), and QSM by at least 4 dB.

The remainder of this paper is organized as follows. In Section II, we introduce the system model and QSM. In Section III, a detailed description of CQSM is given and the rotation angle is numerically optimized and analytically derived for the case of QPSK. A performance evaluation and the computational complexity of CQSM are addressed in Sections IV and V, respectively. In Section VI, simulation results are provided and the convergence of the optimal rotation angle in the case of relatively large-scale systems is addressed. In Section VII, the conclusions are presented.

## II. System Model and Related Work

### 1. System Model

We consider a communication system in which a base station (BS) equipped with $n_T$ transmit antennas communicates on the downlink with a mobile station (MS) equipped with $n_R$ receive antennas. At each channel use, the BS sends $M$ bits to the MS on both the signal constellation symbols and the spatial constellation dimensions. A signal constellation set is denoted by $\Omega$ with cardinality of $|\Omega| = 2^q$, where $q$ denotes the number of bits per signal constellation symbol. The elements $s \in \Omega$ have an average power of one; that is, $\mathbb{E}[s^*s] = 1$. Channel matrix $\mathbf{H}$ couples the $n_R$ receive and $n_T$ transmit antennas, where its element $h_{i,j} \in \mathbb{C}$ is a circularly symmetric complex Gaussian variable with a mean and variance of zero and unity, respectively. The $n_R$-dimensional additive noise vector at the receiver is denoted by $\mathbf{n}$, whose element $n_i$ is circularly symmetric complex Gaussian with a mean and variance of zero and $\sigma_n^2$, respectively.

### 2. Quadrature Spatial Modulation (QSM)

The operation of the QSM scheme is explained through the following numeric example. Assume that the following message $m = [1\ 1\ 0\ 1\ 0\ 0]$, where $|m| = M$, is to be transmitted at a particular channel using QPSK modulation and four transmit antennas. The first two bits $[1\ 1]$ modulate a complex symbol, $s = s_\Re + js_\Im = 1 - j$, where $s_\Re$ and $s_\Im$ are the real and imaginary parts of $s$, respectively. The following $\log_2(n_T)$ bits $[0\ 1]$ modulate the antenna index $l_\Re$ of the antenna used to transmit the real part. The last $\log_2(n_T)$ bits $[0\ 0]$ modulate the index $l_\Im$ of the antenna used to transmit the imaginary part. Therefore, the vectors obtained for real and imaginary parts of the symbol $s$ are given by $\mathbf{s}_\Re = [0\ 1\ 0\ 0]^T$ and $\mathbf{s}_\Im = [-1\ 0\ 0\ 0]^T$, resulting in a transmitted vector given by $\mathbf{s} = \mathbf{s}_\Re + j\mathbf{s}_\Im = [-j\ 1\ 0\ 0]^T$. The transmitted vector is then normalized so that the average power per signal constellation symbol is equal to one. The operator $[.]^T$ denotes a vector/matrix transpose. Based on this description, the spectral efficiency (quantified in bits/s/Hz)

of the QSM scheme is given by:

$$c_{qsm} = q + 2\log_2(n_{n_T}). \quad (1)$$

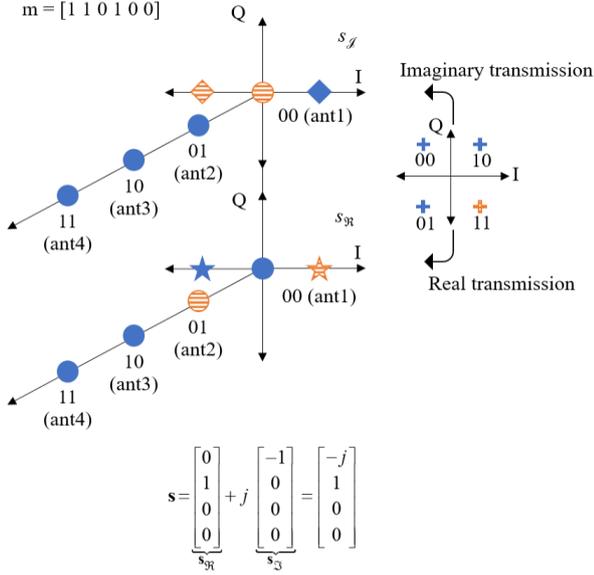

Fig. 1. Example of the QSM scheme using QPSK and four transmit antennas.

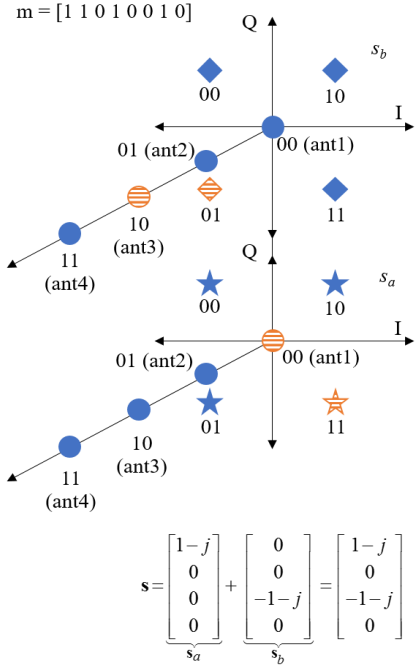

Fig. 2. Example of the CQSM scheme using QPSK and four transmit antennas.

This example is depicted in Fig. 1. The received vector $\mathbf{y} \in \mathbb{C}^{nR \times 1}$ is given by:

$$\mathbf{y} = \mathbf{Hs} + \mathbf{n} = \mathbf{h}_\alpha s_\Re + j\mathbf{h}_\beta s_\Im + \mathbf{n}, \quad \alpha, \beta \in 1, \cdots, n_T \quad (2)$$

where $\alpha$ and $\beta$ are the indices of the spatial constellation symbols, that is, the antennas, from which the real and imaginary parts of $s$ are transmitted, respectively. In addition, $\mathbf{h}_\alpha$ is the $\alpha$-th column of channel matrix $\mathbf{H}$. Based on Fig. 1, the non-zero elements of $\mathbf{s}$ are drawn from either a BPSK alphabet when $\alpha \neq \beta$ or from a QPSK alphabet when $\alpha = \beta$. At the receiver side, the ML detector of the QSM scheme finds $\alpha$, $\beta$, $s_\Re$, and $s_\Im$ as follows:

$$(\hat{\alpha}, \hat{\beta}, \hat{s}_\Re, \hat{s}_\Im) = \underset{\alpha, \beta, s_\Re, s_\Im}{\arg\min} \left\| \mathbf{y} - \mathbf{h}_\alpha s_\Re - j\mathbf{h}_\beta s_\Im \right\|^2. \quad (3)$$

## III. Complex Quadrature Spatial Modulation (CQSM)

The CQSM scheme transmits two complex signal constellation symbols at each channel use. On the other hand, QSM transmits a single complex symbol per channel use. The CQSM transmission leads to a spectral efficiency of

$$c_{cqsm} = 2q + 2\log_2(n_{n_T}). \quad (4)$$

The CQSM scheme is explained through the following numeric example. Assume that message $m = [1\ 1\ 0\ 1\ 0\ 0\ 1\ 0]$ is to be transmitted at a particular channel use by employing QPSK and four transmit antennas. We initially assume that the two symbols are drawn from the same constellation set. It will be later shown that this trivial choice is not optimal. The first and second pair of bits, that is [1 1] and [0 1], modulate the complex symbols, $s_a = (1-j)$ and $s_b = (-1-j)$, respectively. The third and fourth pair of bits, [0 0] and [1 0], modulate the indices of the antennas from which $s_a$ and $s_b$, respectively, are transmitted. The obtained vectors, having the only non-zero elements $s_a$ and $s_b$, are given by $\mathbf{s}_a = [s_a\ 0\ 0\ 0]^T$ and $\mathbf{s}_b = [0\ 0\ s_b\ 0]^T$. Finally, the transmitted vector is given by $\mathbf{s} = \mathbf{s}_a + \mathbf{s}_b$. This example is depicted in Fig. 2.

Based on the CQSM scheme, the received vector $\mathbf{y} \in \mathbb{C}^{nR \times 1}$ is therefore given by

$$\mathbf{y} = \mathbf{Hs} + \mathbf{n} = \mathbf{h}_\alpha s_a + \mathbf{h}_\beta s_b + \mathbf{n}, \quad \alpha, \beta \in 1, \cdots, n_T \quad (5)$$

and the ML receiver for the CQSM is given by

$$\begin{aligned}(\hat{\alpha}, \hat{\beta}, \hat{s}_a, \hat{s}_b) &= \underset{\alpha, \beta, s_a \in \Omega_a, s_b \in \Omega_b}{\arg\min} \left\| \mathbf{y} - (\mathbf{h}_\alpha s_a + \mathbf{h}_\beta s_b) \right\|^2 \\ &= \underset{\alpha, \beta, s_a \in \Omega_a, s_b \in \Omega_b}{\arg\min} \left\| \mathbf{g} \right\|^2 - 2\operatorname{Re}\{\mathbf{y}^H \mathbf{g}\}\end{aligned} \quad (6)$$

where $\mathbf{g} = (\mathbf{h}_\alpha s_a + \mathbf{h}_\beta s_b)$. For the example depicted in Fig. 2, all hypotheses of $\mathbf{h}_\alpha s_a + \mathbf{h}_\beta s_b$ for $\alpha, \beta \in \{1, \ldots, n_T\}$, $s_a \in \Omega_a$ and $s_b \in \Omega_b$ are generated. The hypothesis that minimizes the square Euclidean norm in (6) is the maximum-likelihood (ML) solution. $\alpha$ and $\beta$ of the ML solution are converted to binary, with 1 corresponding to 00, 2 to 01, and so on. The symbols $s_a$ and $s_b$ corresponding to the ML solution are demodulated to the equivalent binary bits, as shown Fig. 2. Accordingly, all the bits of transmitted message $m$ are recovered.

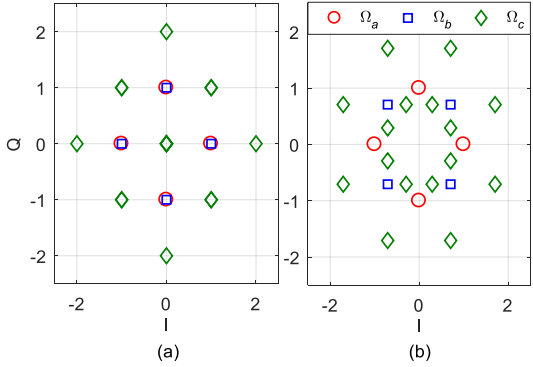

Fig. 3. Example of the received signal constellation set for (a) $\theta = 0$, and (b) $\theta = \pi/4$, using QPSK modulation set $\Omega_a$ and its rotated version $\Omega_b$ at the transmitter side.

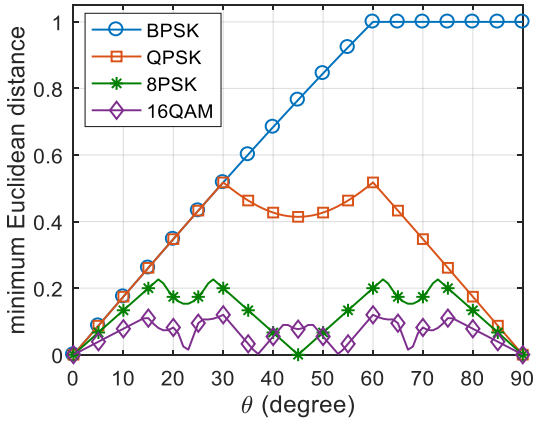

Fig. 4. Minimum Euclidean distance among the members of set $\Omega_d$ versus rotation angle $\theta$ for BPSK, QPSK, 8PSK, and 16QAM modulation sets of $\Omega_a$.

In the sequel, and for the sake of simplicity, we assume the case of QPSK modulation, where a straightforward extension of our conjecture is possible for other cases. Considering (5), there are two distinct cases impacted by the values of $\alpha$ and $\beta$: $\alpha \neq \beta$, and $\alpha = \beta$.

In the first case, where the signal constellation symbols are transmitted from different antennas, the elements of **s** belong to the set $\Omega_d = \Omega_a \cup \Omega_b$. Based on (6), the performance of the ML receiver depends, among other factors, on the minimum Euclidean distance between each pair of symbols in the resulting modulation set, $\Omega_d$, where a larger minimum Euclidean distance results in better performance. To maximize the minimum Euclidean distance between each pair of symbols in the resulting set $\Omega_d$, it suffices to define $\Omega_b$ as:

$$\Omega_b = \{s_b = s_i e^{j\pi/4} \mid s_i \in \Omega_a, i = 1, \cdots, |\Omega_a|\}, \quad (7)$$

where $e^{j\pi/4}$ is a unitary rotation, or, simply, a rotation, with angle of $\pi/4$. This rotation does not change the power of the signal symbols or the angle between them. Therefore, $\Omega_b$ remains a valid QPSK constellation set. In CQSM, rotating the symbols in $\Omega_b$ is required to make the symbols in set $\Omega_a \cup \Omega_b$ unique. Consequently, the signal detection at the receiver side becomes possible. The rotation angle is then optimized so that the bit-error rate is minimized.

On the other hand, in the second case, that is $\alpha = \beta = k$, the system modeled in (5) is rewritten as

$$\mathbf{y} = \mathbf{h}_k(s_a + s_b) + \mathbf{n} = \mathbf{h}_k s_c + \mathbf{n}, \quad k = 1, \cdots, n_T \quad (8)$$

where $s_c \in \Omega_c$, which is defined as

$$\Omega_c = \{s_a + s_b \mid s_a \in \Omega_a, s_b \in \Omega_b, a, b = 1, \cdots, 4\}. \quad (9)$$

By definition, $\Omega_c$ is the Minkowski sum of the sets $\Omega_a$ and $\Omega_b$, which is referred to as $\Omega_c = \Omega_a \oplus \Omega_b$ [17]. Further exploration of $\Omega_c$ in the case of QPSK modulation is given in Appendix I. On one hand, the performance of the ML receiver depends, among other factors, on the minimum distance among the finite lattice points whose basis is matrix **H** working on the transmitted vector **s**, whose non-zero element $s_i \in \Omega_d = \Omega_a \cup \Omega_b \cup \Omega_c$. On the other hand, the BER performance also crucially depends on the minimum Euclidean distance among the signal constellation points, $s_i \in \Omega_d$. At this point, we define the two rotation angles to be optimized in the sequel:

1. $\theta(\mathbf{s})$ is rotation angle $\theta$, such that $s_a \in \Omega_a$ and $s_b \in \Omega_b = s_a e^{j\theta}$. That is, the constellation set $\Omega_b$ is a rotated version of $\Omega_a$. The optimal value of this angle, referred to as $\theta_{opt}(\mathbf{s})$, maximizes the minimum Euclidean distance among the symbols in $\Omega_d$.
2. $\theta(\mathbf{s}, \mathbf{H})$ is the same as $\theta(\mathbf{s})$, except its optimal value, referred to as $\theta_{opt}(\mathbf{s}, \mathbf{H})$, minimizes the BER of the whole system.

First, we address the optimization of $\theta(\mathbf{s})$. The obvious choice of the rotation angle $\theta = \pi/4$ made in the case of $\alpha \neq \beta$ is no longer valid. Figures 3(a) and (b) depict an example of the resulting constellation set $\Omega_d$ for the rotation angles $\theta = 0$ and $\pi/4$, respectively, where the alphabet of each set is indicated with a unique marker.

In Fig. 3(a), the rotation angle $= 0$, which implies that $\Omega_a = \Omega_b$. In the signal constellation set $\Omega_d$, the symbols $0+j0$, $(1+j1)$, $(-1+j)$, $(-1-j)$, $(1-j)$ are repeated four, two, two, two, and two times, respectively, and each element of $\Omega_a$ is repeated twice. The demodulation of the received signal is impossible because the mapping from $\Omega_a$ and $\Omega_b$ to $\Omega_d$ does not result in unique signal modulation symbols. Figure 3(b), on the other hand, depicts the resulting constellation set $\Omega_d$ for $\theta = \pi/4$, where the symbols in the set $\Omega_d$ are unique, making it possible to demodulate the received signal using the ML detector by using (6). Based on this discussion, the choice of $\theta(\mathbf{s})$ should satisfy the following two conditions:

1. The elements of set $\Omega_d$ must be unique. It is therefore

intuitively concluded that, in the case of QPSK modulation, $\theta = 0$ and $\theta = \pi/2$ are to be excluded because these choices result in identical $\Omega_a$ and $\Omega_b$.

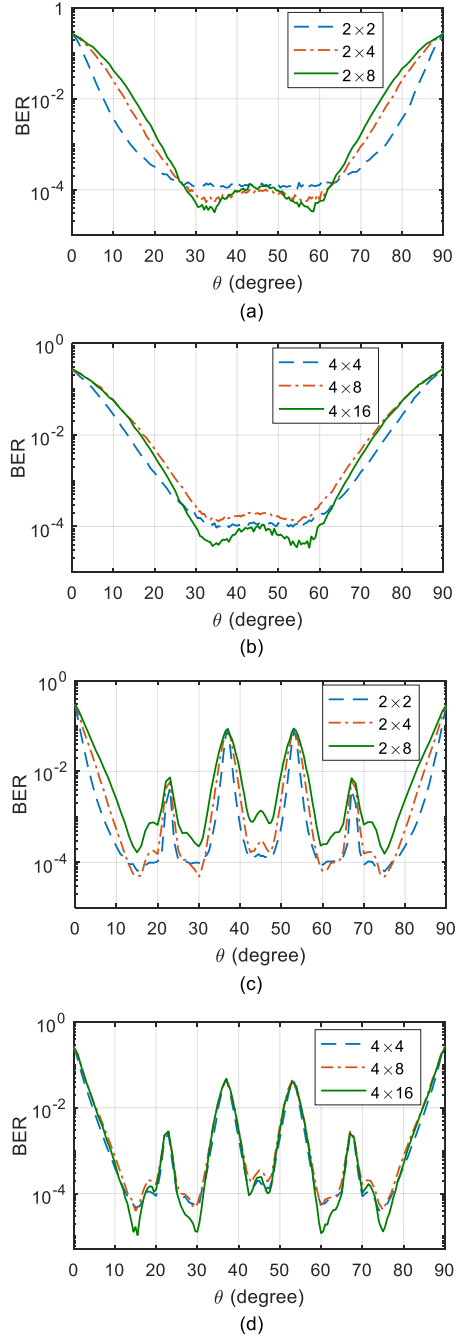

Fig. 5. Minimum Euclidean distance among the members of set $\Omega_d$ versus rotation angle $\theta$ for BPSK, QPSK, 8PSK, and 16QAM modulation sets $\Omega_a$.

2. The minimum Euclidean distance between each pair of symbols in set $\Omega_d$ must be maximized. This is motivated by the fact that the performance of the ML receiver depends on the minimum distance between the received signal constellation points, among other factors.

The optimization problem is therefore given by

$$\hat{\theta}_{opt}(\mathbf{s}) = \arg\max_{\theta \in [0, \pi/2]} d_{\min}(\Omega_d) = \arg\max_{\theta \in [0, \pi/2]} \left( \min_{\substack{s_i, s_k \in \Omega_d \\ i \neq k}} \|s_i - s_k\|^2 \right) \quad (10)$$

Figure 4 depicts the minimum Euclidean distance among the symbols in the constellation set $\Omega_d$, denoted $d_{\min}(\Omega_d)$, for BPSK, QPSK, 8PSK, and 16QAM versus the rotation angle in the range from 0 to $\pi/2$ rad. The results are obtained through simulations. In Appendix I, the optimal rotation angle is analytically obtained in the case of QPSK modulation, where the analytical results and results shown in Fig. 4 coincide. For BPSK, rotation angles greater than or equal to 60° will maximize $d_{\min}(\Omega_d)$. In the case of the other modulation schemes, the curves are even symmetric at approximately $\theta = 45°$. For instance, the case of QPSK with $\theta = \pi/6$ and $\pi/3$ results in a maximum $d_{\min}(\Omega_d)$.

Table 1. Optimal rotation angle(s) $\theta_{opt}(\mathbf{s})$ and corresponding max-min Euclidean distance among the symbols in signal constellation set $\Omega_d$ for different modulation schemes.

| Modulation | $\theta_{opt}(\mathbf{s})$ (degree) | $d_{\min}(\Omega_d)$ |
|---|---|---|
| BPSK | $\geq 60$ | 1 |
| QPSK | 30, 60 | 0.518 |
| 8PSK | 17.3, 27.7, 62.3, 72.7 | 0.230 |
| 16QAM | 30, 60 | 0.119 |

Table 2. Optimal rotation angle(s) $\theta_{opt}(\mathbf{s}, \mathbf{H})$ that minimize(s) the BER rate of the ML receiver for several system configurations.

| $(n_T, n_R)$ | Modulation | $\theta_{opt}(\mathbf{s},\mathbf{H})$ | $(n_T, n_R)$ | Modulation | $\theta_{opt}(\mathbf{s},\mathbf{H})$ |
|---|---|---|---|---|---|
| (2, 2) | QPSK | $\geq 30$ | (4, 5) | QPSK | 35 |
| | 16QAM | 15 | | 16QAM | 15 |
| (2, 4) | QPSK | 32.5 | (4, 6) | QPSK | 35.5 |
| | 16QAM | 15 | | 16QAM | 30.5 |
| (2, 8) | QPSK | 34.5 | (4, 8) | QPSK | 35.5 |
| | 16QAM | 15 | | 16QAM | 30.5 |
| (4, 4) | QPSK | 35 | (8, 8) | QPSK | 38 |
| | 16QAM | 15 | (16,16) | QPSK | 41 |

Table 1 summarizes the optimal rotation angle $\theta_{opt}(\mathbf{s})$ and the corresponding $d_{\min}(\Omega_d)$. Considering Fig. 4 and Table 1, it is worth clarifying that, in the case of 16QAM, $d_{\min}(\Omega_d)$ at $\theta = 30°$ is slightly greater than that at $\theta = 14.7°$. For 8PSK, the distance $d_{\min}(\Omega_d)$ in the range $[0, \pi/4]$ is even symmetric around $\pi/8$ rad.

As stated earlier, there are several variables that affect the BER performance of the ML receiver. Among these variables, rotation angle $\theta$ plays an important role. It is therefore interesting to optimize the rotation angle, taking into consideration all other system parameters, including the channel matrix and system configuration, to mention few. In the sequel, the optimal rotation angle that minimizes the BER performance of the whole system, referred to as $\theta_{opt}(\mathbf{s}, \mathbf{H})$, is optimized. The optimal rotation angle for the ML receiver is obtained through Monte Carlo simulations for QPSK and 16QAM schemes and several $n_T \times n_R$ scenarios. The results are depicted in Fig. 5. Table 2 summarizes these results. The curves depicted in Fig. 5 are simulated for different values of $n_T$ and $n_R$. The SNR value in each scenario is chosen such that the minimum BER is around $10^{-4}$. When the rotation angle leads to a small Euclidean distance among the symbols of $\Omega_d$, for example $\theta = 23º$ and $37º$ in Fig. 5(c), the effect of the rotation angle will dominate that of SNR, leading to almost the same BER performance regardless of the system configuration. However, when the rotation angle leads to a large Euclidean distance among the symbols of $\Omega_d$, for example $\theta = 30º$ in Fig. 5(c), the effect of SNR becomes more obvious.

By comparing Figs. 4 and 5, we conclude that the curves of $d_{min}(\Omega_d)$ versus $\theta(\mathbf{s})$, and BER versus $\theta(\mathbf{s}, \mathbf{H})$, have the same shape; an increase in $d_{min}(\Omega_d)$ leads to improvement in the BER performance. However, the values of $\theta_{opt}(\mathbf{s})$ and $\theta_{opt}(\mathbf{s}, \mathbf{H})$ are different for the same system settings. This is mainly due to the number of used transmit and receive antennas. Further explanation is given in light of the simulation results in Section VI.

## IV. Performance Analysis of CQSM

Let $\mathbf{g} = \mathbf{h}_\alpha s_a + \mathbf{h}_\beta s_b$ and $\hat{\mathbf{g}} = \hat{\mathbf{h}}_\alpha \hat{s}_a + \hat{\mathbf{h}}_\beta \hat{s}_b$ be two received vector signals. Then, the pairwise error probability (PEP) is given as

$$\Pr[\mathbf{g} \to \hat{\mathbf{g}} | \mathbf{H}] = Q\left(\sqrt{\frac{\|\mathbf{g} - \hat{\mathbf{g}}\|^2}{2\sigma_n^2}}\right) = Q\left(\sqrt{\zeta}\right) \quad (11)$$

where

$$\zeta = \frac{1}{2\sigma_n^2}(\mathbf{g} - \hat{\mathbf{g}})^H(\mathbf{g} - \hat{\mathbf{g}}), \quad (12)$$

and its expected value is given in (15), where $\sigma_h^2$ is the variance of the channel gain. The average PEP assuming $n_R$ receive antennas is given by [14], [19]

$$\overline{P}_e(\mathbf{g} \to \hat{\mathbf{g}}) = \gamma^{n_R} \sum_{k=0}^{n_R-1} \binom{n_R-1+k}{k}[1-\gamma]^k \quad (13)$$

where $\gamma = \frac{1}{2}\left(1 - \sqrt{\frac{\overline{\zeta}/2}{1+\overline{\zeta}/2}}\right)$. The BER of the CQSM is upper-bounded by the following average bit-error probability (ABEP):

$$P_b = \frac{1}{2^M}\sum_{i=1}^{2^M}\sum_k^{2^M}\frac{1}{M}\overline{P}_e(\mathbf{g}_i \to \hat{\mathbf{g}}_k)e_{i,k} \quad (14)$$

where $e_{i,k}$ is the number of bit errors associated with $\overline{P}_e(\mathbf{g}_i \to \hat{\mathbf{g}}_k)$.

$$\overline{\zeta} = \begin{cases}
\frac{\sigma_h^2}{2\sigma_n^2}(|x_a|^2 + |x_b|^2 + |\hat{x}_a|^2 + |\hat{x}_b|^2) \\
\quad \text{if } \mathbf{h}_\alpha \neq \mathbf{h}_\beta, \hat{\mathbf{h}}_\alpha \neq \hat{\mathbf{h}}_\beta, \mathbf{h}_\alpha \neq \hat{\mathbf{h}}_\alpha, \mathbf{h}_\beta \neq \hat{\mathbf{h}}_\beta \\
\frac{\sigma_h^2}{2\sigma_n^2}(|x_a - \hat{x}_a|^2 + |x_b|^2 + |\hat{x}_b|^2) \\
\quad \text{if } \mathbf{h}_\alpha \neq \mathbf{h}_\beta, \hat{\mathbf{h}}_\alpha \neq \hat{\mathbf{h}}_\beta, \mathbf{h}_\alpha = \hat{\mathbf{h}}_\alpha, \mathbf{h}_\beta \neq \hat{\mathbf{h}}_\beta \\
\frac{\sigma_h^2}{2\sigma_n^2}(|x_b - \hat{x}_b|^2 + |x_a|^2 + |\hat{x}_a|^2) \\
\quad \text{if } \mathbf{h}_\alpha \neq \mathbf{h}_\beta, \hat{\mathbf{h}}_\alpha \neq \hat{\mathbf{h}}_\beta, \mathbf{h}_\alpha \neq \hat{\mathbf{h}}_\alpha, \mathbf{h}_\beta = \hat{\mathbf{h}}_\beta \\
\frac{\sigma_h^2}{2\sigma_n^2}(|x_a - \hat{x}_a|^2 + |x_b - \hat{x}_b|^2) \\
\quad \text{if } \mathbf{h}_\alpha \neq \mathbf{h}_\beta, \hat{\mathbf{h}}_\alpha \neq \hat{\mathbf{h}}_\beta, \mathbf{h}_\alpha = \hat{\mathbf{h}}_\alpha, \mathbf{h}_\beta = \hat{\mathbf{h}}_\beta \\
\frac{\sigma_h^2}{2\sigma_n^2}(|x_a + x_b|^2 + |\hat{x}_a|^2 + |\hat{x}_b|^2) \\
\quad \text{if } \mathbf{h}_\alpha = \mathbf{h}_\beta, \hat{\mathbf{h}}_\alpha \neq \hat{\mathbf{h}}_\beta, \mathbf{h}_\alpha \neq \hat{\mathbf{h}}_\alpha, \mathbf{h}_\beta \neq \hat{\mathbf{h}}_\beta \\
\frac{\sigma_h^2}{2\sigma_n^2}(|x_a + x_b - \hat{x}_a|^2 + |\hat{x}_b|^2) \\
\quad \text{if } \mathbf{h}_\alpha = \mathbf{h}_\beta, \hat{\mathbf{h}}_\alpha \neq \hat{\mathbf{h}}_\beta, \mathbf{h}_\alpha = \hat{\mathbf{h}}_\alpha, \mathbf{h}_\beta \neq \hat{\mathbf{h}}_\beta \\
\frac{\sigma_h^2}{2\sigma_n^2}(|x_a + x_b - \hat{x}_b|^2 + |\hat{x}_a|^2) \\
\quad \text{if } \mathbf{h}_\alpha = \mathbf{h}_\beta, \hat{\mathbf{h}}_\alpha \neq \hat{\mathbf{h}}_\beta, \mathbf{h}_\alpha \neq \hat{\mathbf{h}}_\alpha, \mathbf{h}_\beta = \hat{\mathbf{h}}_\beta \\
\frac{\sigma_h^2}{2\sigma_n^2}(|\hat{x}_a + \hat{x}_b|^2 + |x_a|^2 + |x_b|^2) \\
\quad \text{if } \mathbf{h}_\alpha \neq \mathbf{h}_\beta, \hat{\mathbf{h}}_\alpha = \hat{\mathbf{h}}_\beta, \mathbf{h}_\alpha \neq \hat{\mathbf{h}}_\alpha, \mathbf{h}_\beta \neq \hat{\mathbf{h}}_\beta \\
\frac{\sigma_h^2}{2\sigma_n^2}(|x_a - \hat{x}_a - \hat{x}_b|^2 + |x_b|^2) \\
\quad \text{if } \mathbf{h}_\alpha \neq \mathbf{h}_\beta, \hat{\mathbf{h}}_\alpha = \hat{\mathbf{h}}_\beta, \mathbf{h}_\alpha = \hat{\mathbf{h}}_\alpha, \mathbf{h}_\beta \neq \hat{\mathbf{h}}_\beta \\
\frac{\sigma_h^2}{2\sigma_n^2}(|x_b - \hat{x}_a - \hat{x}_b|^2 + |x_a|^2) \\
\quad \text{if } \mathbf{h}_\alpha \neq \mathbf{h}_\beta, \hat{\mathbf{h}}_\alpha = \hat{\mathbf{h}}_\beta, \mathbf{h}_\alpha \neq \hat{\mathbf{h}}_\alpha, \mathbf{h}_\beta = \hat{\mathbf{h}}_\beta \\
\frac{\sigma_h^2}{2\sigma_n^2}(|x_a + x_b|^2 + |\hat{x}_a + \hat{x}_b|^2) \\
\quad \text{if } \mathbf{h}_\alpha = \mathbf{h}_\beta, \hat{\mathbf{h}}_\alpha = \hat{\mathbf{h}}_\beta, \mathbf{h}_\alpha \neq \hat{\mathbf{h}}_\alpha, \mathbf{h}_\beta \neq \hat{\mathbf{h}}_\beta \\
\frac{\sigma_h^2}{2\sigma_n^2}(|x_a + x_b - \hat{x}_a - \hat{x}_b|^2) \\
\quad \text{if } \mathbf{h}_\alpha = \mathbf{h}_\beta, \hat{\mathbf{h}}_\alpha = \hat{\mathbf{h}}_\beta, \mathbf{h}_\alpha = \hat{\mathbf{h}}_\alpha, \mathbf{h}_\beta = \hat{\mathbf{h}}_\beta
\end{cases} \quad (15)$$

## V. Computational Complexity

The evaluation of the second line of (6) requires ($4n_R+1$) real multiplications and ($4n_R-1$) real additions. Since the ML search is performed over an $M$-dimensional space, CQSM requires the following number of real multiplications and additions:

$$\eta^{\text{mul}} = (4n_R + 1)2^M,$$
$$\eta^{\text{add}} = (4n_R - 1)2^M. \quad (11)$$

Since the ML detectors for both CQSM and QSM have the same form of optimization function, given above in (6) and (3) in [14], respectively, both schemes have equivalent computational complexity for the same spectral efficiency of $M$ bits/s/Hz. The computational complexity required to evaluate (6) is listed in Table 3.

Table 3. Receiver computational complexity.

| Term | Real multiplications | Real additions |
|---|---|---|
| $\|\mathbf{g}\|^2$ | $2n_R$ | $2n_R-1$ |
| $2\text{Re}\{\mathbf{y}^H\mathbf{g}\}$ | $2n_R+1$ | $2n_R-1$ |
| $\|\mathbf{g}\|^2-2\text{Re}\{\mathbf{y}^H\mathbf{g}\}$ | 0 | 1 |
| Total | $4n_R+1$ | $4n_R-1$ |

## VI. Simulation Results and Discussion

In this section, the receiver is considered to have perfect knowledge of the channel state information. In addition, data bits are considered to be random such that the signal and spatial symbols are uniformly distributed. The optimal rotation angles used to obtain the simulation results are depicted in Table 2.

Figure 6 shows the BER performance of SM, GSM, QSM and CQSM for the same spectral efficiency of 8 bits/s/Hz. The GSM transmitter is equipped with $n_T = 7$ and employs a combination of $n_U = 2$ antennas to transmit one signal symbol at each channel use. Assuming both SM and GSM use 16QAM, SM requires 16 transmit antennas to achieve the same spectral efficiency, compared to only seven used by GSM. The reduction in $n_T$ comes at a moderate computational cost [18]. When SM uses 64QAM and four transmit antennas, it lags the performance of GSM by approximately 3 dB at a target BER of $10^{-4}$. Therefore, a tradeoff between performance and the number of transmit antennas can be achieved.

On the other hand, CQSM outperforms SM and QSM by 7.1 dB and 5.1 dB, respectively, at a target BER of $10^{-4}$. Moreover, CQSM outperforms GSM by 4.5 dB while requiring four transmit antennas, compared to seven required by GSM. Finally, CQSM outperforms QSM by 5.1 dB in the case of the 4×8 system. These enhancements of CQSM incur no additional computational costs, as detailed in Section V.

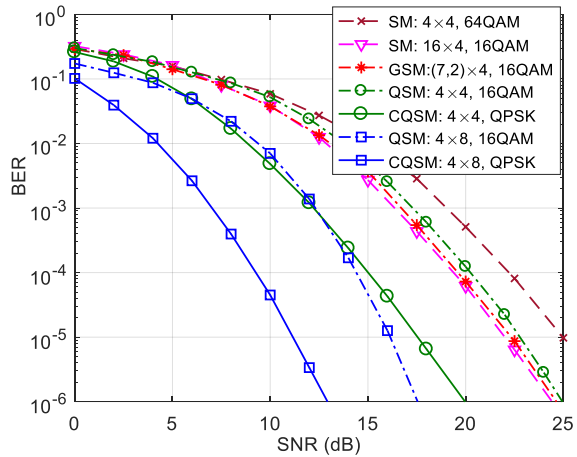

Fig. 6. BER performances of SM, GSM, QSM and CQSM schemes for the same spectral efficiency of 8 bits/s/Hz.

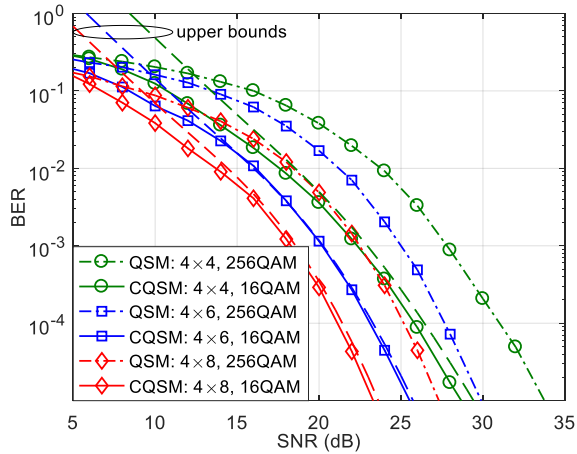

Fig. 7. BER performance of QSM and CQSM for the same transmission rate of 12 bits/s/Hz.

Figure 7 depicts the performance of CQSM and QSM schemes using 16QAM and 256QAM, respectively, where both schemes achieve the same transmission rate. The upper bound of the ABEP of CQSM given in (14) is also shown for the considered scenarios. CQSM still outperforms the QSM scheme by 4.1, 4.5, and 5.2 dB in the case of 4×8, 4×6, and 4×4 systems, respectively. The outperformance of CQSM is slightly reduced as the number of receive antennas is increased.

Figure 8 depicts the performance of CQSM and QSM schemes for data rates of 8 and 6 bits/s/Hz, respectively, using QPSK modulation. The analytical ABEP of CQSM given in (14) is also shown for the considered scenarios. QSM outperforms CQSM by 0.5, 0.57, and 1 dB for 4×4, 4×6, and 4×8 systems, respectively. This degradation is tolerable as the CQSM scheme increases the achieved spectral efficiency by

33.33%.

Finally, the performance of CQSM and QSM schemes for several $n_T = n_R$ scenarios using QPSK modulation are depicted in Fig. 9. The proposed system achieves an increase of 50%, 33.33%, 25% and 20% in spectral efficiency using 2, 4, 8, and 16 transmit antennas, respectively. As the number of transmit antennas increases, the performance gap between CQSM and QSM decreases. In the case of $n_T = 16$, the proposed scheme outperforms QSM for most of the simulated values of SNR. This performance trend is explained in the sequel. Without loss of generality, we consider the case of QPSK modulation because the following conjecture can be simply extended to any other modulation set.

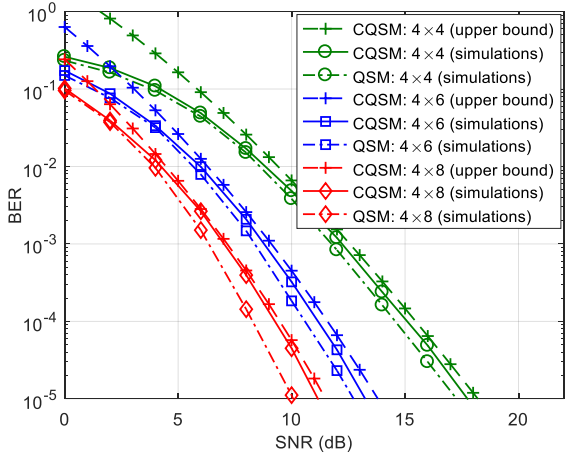

Fig. 8. BER performance of QSM and CQSM with transmission rates of 6 and 8 bits/s/Hz, respectively, using QPSK modulation for several system configurations.

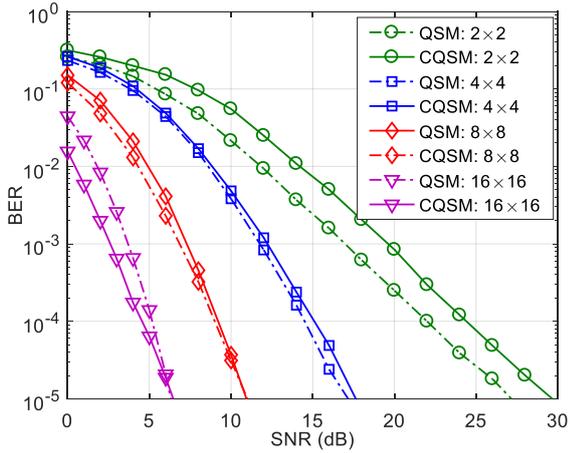

Fig. 9. BER performance of QSM and CQSM using QPSK modulation. CQSM achieves 6, 8, 10, and 12 bits/s/Hz versus 4, 6, 8, 10 bits/s/Hz for QSM, using 2, 4, 8, and 16 transmit antennas, respectively.

Let $s_i$ be a non-zero element of the transmitted vector **s**. Then, the following two probabilities hold true.

$$\Pr[s_i \in \Omega_c] = \frac{1}{n_T},$$

$$\Pr[s_i \in \Omega_a \mid s_i \in \Omega_b] = \frac{n_T - 1}{n_T}.$$

As $n_T$ grows large, the following two limits are satisfied:

$$\lim_{n_T \to \infty} \Pr[s_i \in \Omega_c] = 0,$$

$$\lim_{n_T \to \infty} \Pr[s_i \in \Omega_a \mid s_i \in \Omega_b] = 1.$$

This implies that, at a very large $n_T$, $\Omega_d \approx \Omega_a \cup \Omega_b$. In this case, the optimum rotation angle is intuitively given by $\theta_{opt}(\mathbf{s}) = \theta_{opt}(\mathbf{s}, \mathbf{H}) = 45°$ and $d_{min}(\Omega_d) = 2\sin(\theta/2) = 0.765$. This value is larger than 0.518; the Euclidean distance at $\theta = 30°$ when the symbols in $\Omega_d$ are considered to be equally probable. This conjecture coincides with the optimal values of the rotation angle listed in Table 2: $\theta_{opt}(\mathbf{s}, \mathbf{H})$ increases from 30 to 41 when the number of transmit antennas $n_T$ increases from 2 to 16. Using Monte Carlo simulations, it is additionally determined that the optimal rotation angle $\theta_{opt}(\mathbf{s}, \mathbf{H})$ obtained for the 32×32 system is 41.5°. This result coincides with our conjecture on the convergence of the optimal rotation angle presented in Section III.

## VII. Conclusions

In this paper, we proposed a CQSM scheme, where two complex constellation symbols drawn from two different modulation sets are transmitted at each channel use, leading to a higher transmission rate compared to QSM. The first symbol is drawn from a conventional QAM/PSK modulation set; the second is drawn from a rotated version of the former set. Since the rotation angle affects the system BER performance, it was optimized using Monte Carlo simulations as well as analytically. Simulation results showed that, for the same transmission rate, CQSM outperformed the GSM and QSM by at least 4 to 5 dB in several system settings. It was also numerically and analytically shown that, as the number of transmit antennas became large, CQSM outperformed QSM while achieving a higher transmission rate.

## Appendix I

Let $\Omega_a$ be given by

$$\Omega_a = \{s_i = e^{j(i-1)\pi/2} \mid i = 1, \cdots, 4\}$$

Then, the rotated constellation set is given by

$$\Omega_b = \{s_i = e^{j((i-1)\pi/2 + \theta)} \mid i = 1, \cdots, 4\}$$

where $\theta$ is the rotation angle of the signal constellation set. Based on CQSM, $\Omega_c$ is defined as

$$\Omega_c = \{s_i + s_k \mid s_i \in \Omega_a, s_k \in \Omega_b, i, k = 1, \cdots, 4\}$$

where $\Omega_c = \Omega_a \oplus \Omega_b$ is the Minkowski sum of the sets $\Omega_a$ and $\Omega_b$. From a computational geometry perspective, the Minkowski sum is represented as the union of the following subsets:

$$\Omega_c = \bigcup_{i=1}^{4} \Omega_i$$

where

$$\Omega_i = \{s_i + s_k \mid s_i \in \Omega_b, s_k \in \Omega_a, k = 1, \cdots, 4\}$$

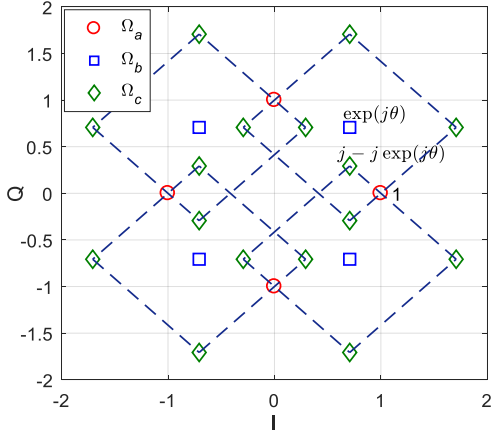

Fig. 10. Example of the resulting signal modulation sets, $\Omega_a$, the rotated set $\Omega_b$, and their Minkowski sum $\Omega_c$ for QPSK modulation and rotation angle $\theta = \pi/4$.

That is, since the symbols in $\Omega_a$ are located at the corners of a square centered at the origin, the subset $\Omega_i$ is a shifted version of $\Omega_a$ and is centered at $s_i \in \Omega_b$. Figure 10 depicts an example of the resulting signal modulation set. It is worth mentioning that:

1. Owing to the structure of the QPSK modulation set, the resulting $\Omega_d$ is even symmetric around $\pi/4$. That is, the values of $d_{\min}(\Omega_d)$ are identical for the angles $\theta$ and $(\pi/2 - \theta)$. Therefore, the optimization of $\theta$ is carried out in the interval $[0, \pi/4]$. Figure 11 depicts $\Omega_d$ for rotation angles $\theta \in [0, \pi/4]$.

2. On account of the symmetry of the resulting signal modulation set $\Omega_d$, the search of the optimum rotation angle reduces to:

$$\theta_{opt}(\mathbf{s}) = \arg\max_{\theta \in [0, \pi/4]} \left( \min(d_1, d_2) \right).$$

Let $s_1 \in \Omega_a = 1$, $s_2 \in \Omega_c = (j - je^{j\theta})$, and $s_3 \in \Omega_b = e^{j\theta}$, as indicated in Fig. 10. Then,

$$d_1 = \|s_1 - s_2\| = \sqrt{3 - 2\sin(\theta) - 2\cos(\theta)}$$

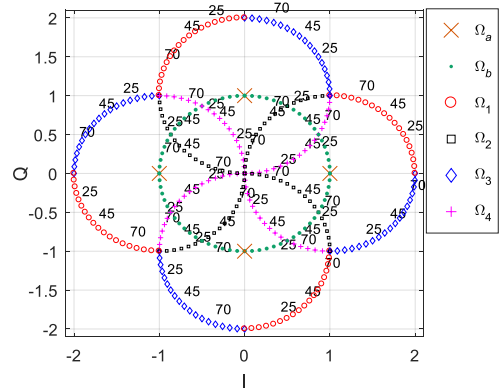

Fig. 11. Signal constellation symbols in $\Omega_d$ for QPSK modulation and rotation angle $\theta \in [0, \pi/2]$.

$$d_2 = \|s_1 - s_3\| = \sqrt{2 - 2\cos(\theta)}$$

Since $d_1$ is a strictly decreasing function and $d_2$ is a strictly increasing function, the optimal rotation angle therefore satisfies the condition $d_1 = d_2$. From the above definitions of $d_1$ and $d_2$, the optimal rotation angle satisfies $\sin(\theta) = 0.5$, which implies that

$$\theta_{opt}(\mathbf{s}) = \frac{\pi}{6}$$

in the case of QPSK modulation.

A second geometrical analysis of $\Omega_c$ that engenders further insight on the resulting receiver performance is given in the sequel. Let $\Omega_c$ be rewritten as a union of four disjoint subsets $\Omega_i$, $i = 1, \ldots, 4$. The symbols of each subset are located at the four corners of a square for any rotation angle $\theta$. Additionally, the symbols belonging to the same subset have the same power and an ensemble mean of zero. Let the polar representation of the first element of the subset $\Omega_i$ be written as $r_i e^{j\phi_i}$. Then, with the contribution of trigonometric analysis, it is shown that

$$r_1 = \sqrt{2 + 2\cos(\theta)}, \quad \phi_1 = \frac{\theta}{2}$$

$$r_2 = \sqrt{2 - 2\sin(\theta)}, \quad \phi_2 = \frac{\theta}{2} + \frac{\pi}{4}$$

$$r_3 = \sqrt{2 + 2\sin(\theta)}, \quad \phi_3 = \frac{\theta}{2} - \frac{\pi}{4}$$

$$r_4 = \sqrt{2 - 2\cos(\theta)}, \quad \phi_4 = \frac{\theta}{2} - \frac{\pi}{2}$$

For instance,

$$s_1 \in \Omega_1 = 1 + e^{j\theta} = 1 + \cos(\theta) + j\sin(\theta).$$

Therefore, $r_1$ is simply given as above, and angle $\phi_1$ is given by

$$\phi_1 = \tan^{-1}\left(\frac{\sin(\theta)}{1+\cos(\theta)}\right) = \tan^{-1}\left(\frac{2\sin(\theta/2)\cos(\theta/2)}{2\cos^2(\theta/2)}\right)$$

$$= \tan^{-1}\left(\frac{\sin(\theta/2)}{\cos(\theta/2)}\right) = \frac{\theta}{2}$$

Figure 11 depicts the sets $\Omega_a$, $\Omega_b$, $\Omega_1$, $\Omega_2$, $\Omega_3$, and $\Omega_4$ for $\theta \in [0, \pi/2]$. The values of several rotation angles are also added to indicate the direction of rotation. These results can be used for the derivation of the optimal rotation angle. Furthermore, they can foster further insight in analyzing the CQSM scheme in the case of high-order PSK modulation.

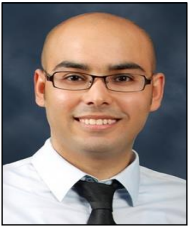
**Manar Mohaisen** received an MSc degree in communications and signal processing from the University of Nice-Sophia Antiplois, France, in 2005, and a PhD from Inha University, Rep. of Korea, in 2010, both in communications engineering. From 2001 to 2004, he was a cell planning engineer at the Palestinian Telecommunications Company. Since 2010, he has been an assistant professor at the Department of EEC Engineering, KoreaTech, Rep. of Korea. His research interests include MIMO systems, communication systems, and social network analysis.

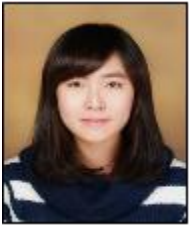
**Saetbyeol Lee** received a BEng degree in electronics engineering from Korea Tech, Rep. of Korea in 2015. She is currently pursuing an MSc degree at the Department of Electrical, Electronics, and Communications Engineering, Korea Tech, Rep. of Korea. Her research interests include MIMO systems with an emphasis on spatial modulation and interference alignment.